\documentclass[12pt]{article}
\usepackage{color}
\usepackage{latexsym}
\usepackage{amsmath}
\usepackage{amsfonts}
\usepackage{amssymb}
\usepackage{indentfirst} 
\numberwithin{equation}{section}
\newcommand{\be}{\begin{equation}}
\newcommand{\ee}{\end{equation}}
 
\newcommand{\mR}{{\mathbb R}}

\title{Nonlocal dynamics and infinite non-relativistic conformal symmetries }
\author{K. Andrzejewski\footnote{E-mail: k-andrzejewski@uni.lodz.pl}
\\
\small Department of  Computer Science, \\
\small University of \L\'od\'z,\\
\small Pomorska 149/153, 90-236 {\L}\'od\'z, Poland\\
\\
K. Bolonek-Laso\'n \footnote{E-mail: kbolonek1@wp.pl}\\
\small Department of Statistical Methods, \\
\small University of \L\'od\'z,\\
\small Rewolucji 1905 St. 41/43, 90-214, {\L}\'od\'z, Poland\\
}
\date{}
\begin{document}
\maketitle 
\begin{abstract}
We study  the symmetry of the class of nonlocal models  which includes the  nonlocal extension of the Pais-Uhlenbeck oscillator. As a consequence,  we obtain an   infinite dimensional symmetry  algebra, containing the Virasoro algebra, which   can be considered as a generalization of the  non-relativistic  conformal symmetries  to  the infinite order. Moreover, 
this nonlocal extension  resembles to some extent the string model and  on the quantum level it leads  to the centrally extended   Virasoro algebra. 
\end{abstract}
\section{Introduction}
Because of the importance of  the noncommutative field  theories  and the string field   theory  the nonlocal models    have been extensively investigated in last two decades (see, e.g.,  \cite{b1a}-\cite{b1g}, as a very brief list of  references). However, the  nonlocality  in physics has a long history  (see \cite{b2} and the references in \cite{b3a}). The  nonlocal systems  provide an  extension  of the  higher derivatives ones to the case of the infinite order  derivatives.  This  fact is   reflected  in   form of the Hamiltonian formalism for the nonlocal theories   which    is a generalization of  the Ostrogradski approach to the  higher derivatives theory \cite{b3a}-\cite{b3e}. 
\par 
  The basic example of the theory with higher derivatives is the  Pais-Uhlenbeck (PU) oscillator of  order $N$ (which generalizes the standard Lagrangian  of the  harmonic oscillator to the case of the $N$-th order time derivative) proposed in the classical paper \cite{b2}.  This model has attracted  interest thought the years (for the last  few years see, e.g.,  \cite{b4}-\cite{b5}).   In particular,  it has been shown  (see  \cite{b5}) that   when   the relevant  eigenfrequencies  are proportional to consecutive odd integers then the maximal symmetry group of the  PU oscillator  is the   $l$-conformal Newton-Hooke group with half-integer $l=N -\frac 12 $, which is  isomorphic to the $l$-conformal Galilei group (for this reason we will simply refer  to it as  the $l$-conformal non-relativistic group in what follows). These groups are natural   conformal extensions of the Newton-Hooke (Galilei) group, (in particular for $N=1$ we obtain the  Schr\"odinger one)  and originally  have been introduced in the context of non-relativistic gravity \cite{b6b,b6c}; however,  there are many dynamical  models,  both classical  (including also  the second order dynamical equations)  and quantum, which exhibit such  symmetries \cite{b6d}-\cite{b6n}.  Furthermore, in a number of papers \cite{b7b}-\cite{b7i} infinite dimensional extensions of  those symmetries, for fixed order $l$, have been also considered in the context of   some more sophisticated models. Let us note that neither the  $l$-conformal non-relativistic algebra nor its infinite extension are well defined  as $N (\textrm{or } l)$ tends to  infinity.    
\par
In the present paper we show that a very  simple nonlocal theory (proposed in ref. \cite{b2}) enjoys   symmetry which can be considered as an infinite-dimensional generalization   of the non-relativistic conformal symmetries to the case of  $N$ going out  to  infinity. The starting point is the mentioned observation that the  $(N-\frac 12)$-conformal  non-relativistic  group appears as a symmetry of  the PU oscillator of order $N$. Next, we note that in ref. \cite{b2}  the authors  introduce (however, slightly  formally)   the $N=  \infty$ extension of the PU oscillator which,  in the case of odd frequencies,  leads  after some manipulations to the  quadratic nonlocal Lagrangian (see also \cite{b3e}).  We find   the symmetry algebra of this nonlocal Lagrangian  and its   central extension  on  the Hamiltonian level.  The symmetry algebra obtained in this way contains two representations of  the Virasoro algebra. In the case  of many spatial dimensions we obtain the Virasoro-Kac-Mody   type algebra. Finally, we show that on the quantum level   the Virasoro generators do  form the centrally extended algebra. 
\par The paper is organized as follows. In Section 2 we describe in some detail a class of nonlocal  Lagrangians and derive the explicit  relation, on the Lagrangian level,   between the symmetry and the  integral of motion.    
In Section  3 we briefly recall the  Hamiltonian formalism for the nonlocal  PU  model; in the next section, we find the  symmetries and corresponding   integrals of motion for this model in  the  Lagrangian  framework. Section 5 is devoted to  the symmetry  realizations on the Hamiltonian and quantum levels with special emphasis put on the Virasoro algebra. Finally,  Section 6 contains the final discussion and sketches the possible ways of further  development.   
\section{Nonlocal systems}
Let us  consider  the following class of nonlocal Lagrangians 
\be
\label{e1}
L=L(q(t),q(t+\alpha)), \quad \alpha>0.
\ee
The action principle
\be
\label{e2}
\delta S\equiv \int_{-\infty}^{\infty} \frac{\delta L(s)}{\delta q(t)}ds=0,
\ee
applied to the Lagrangian (\ref{e1})  implies the following  equation of motion 
\be
\label{e3}
\frac{\partial L(q(t),q(t+\alpha))}{\partial q(t)}+\frac{\partial L(q(t-\alpha),q(t))}{\partial q(t)}=0.
\ee
In order to better  understand  the structure of  the nonlocal theory and to derive the Hamiltonian formalism (and, eventually, the quantum theory)  let us  
expand the Lagrangian (\ref{e1})  in powers of the parameter $\alpha$; then $L$ becomes  a function of all time derivatives of $q(t)$.  Thus it requires an  infinite number of  initial conditions or the knowledge of   a finite piece of the trajectory. Consequently,  we need a Hamiltonian  formalism which would be a generalization of the Ostrogradski one  which is valid only for a finite order  derivatives.  Such  formalism has been proposed  and discussed   in refs.  \cite{b3a}-\cite{b3e}.  In this approach we consider, instead of the trajectory $q(t)$,  $1+1$ dimensional  field  $Q(t,\lambda)$  such that
\be
\label{e4}
Q(t,\lambda)=q(t+\lambda).
\ee  
Our Lagrangian $L$ is rewritten  as follows
\be
\label{e5}
\tilde L(Q(t,0),Q(t,\alpha))= \int _{-\infty}^\infty d\lambda\delta(\lambda)\tilde L(Q(t,\lambda),Q(t,\lambda+\alpha)),
\ee
where $\tilde L(Q(t,\lambda),Q(t,\lambda+\alpha))$ is obtained from the original Lagrangian (\ref{e1}) by  making the following replacements 
\be
\label{e6}
q(t)\rightarrow Q(t,\lambda),\quad q(t+\alpha)\rightarrow Q(t,\lambda+\alpha) .
\ee
Now,  the  Hamiltonian for the field $Q$ is provided by
\be
\label{e7}
H=\int_{-\infty}^\infty d \lambda P(t,\lambda)Q'(t,\lambda)-\tilde L,
\ee
where $P$ is the canonical momentum of $Q$ and  prime  stands for $\partial_\lambda$. The Poisson bracket is of the form
\be
\label{e8}
\{Q(t,\lambda),P(t,\lambda')\}=\delta(\lambda-\lambda ').
\ee
 In order to recover the original dynamics one has to impose new constraint:
\be
\label{e9}
\phi\equiv P(t,\lambda)-\frac{1}{2}\int_{-\infty}^\infty d\sigma(\textrm{sgn}(\lambda)-\textrm{sign}(\sigma) )E(t,\sigma,\lambda)\approx 0,
\ee
where 
\be
\label{e10}
E(t,\sigma,\lambda)=\frac{\delta L(Q(t,\sigma),Q(t,\sigma+\alpha))}{\delta Q(t,\lambda)}.
\ee
The constraint $\phi$ implies the  secondary constrains which can be written collectively  in the following form 
\be
\label{e11}
\psi \equiv \int _{-\infty}^\infty d\sigma E(t,\sigma,\lambda)\approx 0.
\ee 
The constraint (\ref{e11})  is equivalent  to the Euler-Lagrange  equation for the nonlocal Lagrangian (\ref{e1}) and together with  the constraint (\ref{e9}) they  form the second class constraints. So  we can use the Dirac method for constrained systems  and  solve  $\phi$ for $P$. Then the   Hamiltonian 
\begin{align}
\label{e12}
H=&-\frac{1}{2}\int_{-\infty}^\infty d \lambda (\textrm{sgn}(\lambda-\alpha)-\textrm{sign}(\lambda) )\frac{\partial L(Q(t,\lambda-\alpha),Q(t,\lambda))}{\partial Q(t,\lambda)}Q'(t,\lambda)\nonumber \\
& -L(Q(t,0),Q(t,\alpha)),
\end{align}
is expressed only in terms of $Q$  which is subjected to the  single  condition (\ref{e11}) and satisfies commutation relations following from the relevant  Dirac bracket.   
\par
Apart from  the Hamiltonian we can look for another integrals of motion (and the corresponding symmetries)  related to the Lagrangian (\ref{e1}).  
In order to do this let us consider the infinitesimal transformation of the time $t$  and the coordinate $q$
 \be
 \label{e13}
 t'=t+\delta t(t), \quad q'(t')=q(t)+\delta q(t),
 \ee
 satisfying the following condition
 \be
 \label{e14}
 \delta t(t+\alpha)=\delta t(t).
 \ee
 Then 
\be
\label{e15}
t'+\alpha=t+\alpha+\delta t(t+\alpha), \quad
q'(t'+\alpha)=q(t+\alpha)+\delta q(t+\alpha).
\ee 
 Next, let us define
 \be
 \label{e15a}
 \delta_0q(t)=\delta q(t)-\dot q(t)\delta (t),\quad \delta _0q(t+\alpha)=\delta q(t+\alpha)-\dot q(t+\alpha)\delta t(t+\alpha).
 \ee
 The action principle is invariant under the transformations (\ref{e13}) provided  
\be
\label{e16}
L(q'(t'),q'(t'+\alpha))\frac{dt'}{dt}=L(q(t),q(t+\alpha))+\frac{dF}{dt},
\ee
which infinitesimally  takes the form  
\be 
\label{e17}
\frac{d}{dt}(L\delta t-\delta F)+\frac{\partial  L(q(t),q(t+\alpha))}{\partial q(t)}\delta_0 q(t)+\frac{\partial L(q(t),q(t+\alpha ))}{\partial q(t+\alpha)}\delta_0q(t+\alpha)=0.
\ee
Using  eq. (\ref{e17})  one can check, by direct calculations,  that the function $C$ defined as follows 
\be
\label{e18}
C= L\delta t+\int_{t}^{t+\alpha}d \lambda\frac{\partial L(q(\lambda -\alpha),q(\lambda))}{\partial q (\lambda)} \delta _0q(\lambda) -\delta F,
\ee
is an integral of motion associated with the symmetry (\ref{e13}). This result can be also  obtained  following ref.  \cite{b3c} directly within the  Hamiltonian framework keeping in mind  that the constraint (\ref{e9}), being  of second kind, can be used to eliminate the momentum variable $P(t,\lambda)$. Let us note that in the case of the  time translation, i.e., 
\be
\label{e19}
\delta t=-\epsilon, \quad \delta q=0,
\ee
$C$ coincides  with $H$ after making the replacement (\ref{e6})
\section{Nonlocal Pais-Uhlenbeck model}
The PU oscillator of  order $N$  
\be
\label{e19a}
L_{PU}=-\frac{1}{2}  q\prod_{k=1}^{N}\left(\frac{d^2}{dt^2}+\omega_k^2\right)  q, 
\ee
studied  in the original  paper \cite{b2} is  considered there  as a starting point towards the  nonlocal theory. Namely, it has been  shown (rather formally)  that taking   the limit $N\rightarrow\infty$ in (\ref{e19a})  we obtain,  in the case of the odd frequencies $\omega_k=(2k-1)\frac{\pi}{2\alpha}$, $k=1,2,\ldots$,  a  quadratic nonlocal Lagrangian which  (after some  manipulations) can be written in the form   
\be
\label{e20}
L=-\frac{m}{\alpha^2}q(t)q(t+\alpha),
\ee
where $m$ and $\alpha$ are some constants  of  the dimension of mass and time, respectively.
Equation of motion  (\ref{e3}) in this case  yields
\be
\label{e21}
q(t-\alpha)+q(t+\alpha)=0.
\ee
The slightly surprising  interpretation of the Lagrangian (\ref{e20})  as a generalization   of the  PU oscillator with   the  odd frequencies  to the case of the infinite  time derivatives (or, equivalently, with  infinite number of oscillators)  has been recently   confirmed, in two ways,   in ref. \cite{b3e}.
First,  it was shown that,  expanding the right-hand side of (\ref{e20}) in  a Taylor series in $\alpha$ and then restricting ourselves to the first $k$  terms we obtain a  Lagrangian, which  is proportional to the
PU oscillator with some frequencies which  tends to the odd ones in the   limit $k\rightarrow \infty$. The authors of ref. \cite{b3e} applied also the Hamiltonian formalism  discussed in  the previous section and  arrive at the following Hamiltonian dynamics
\be
\label{e22}
H=\frac{m}{2\alpha^2}\int_{-\infty}^\infty d \lambda (\textrm{sgn}(\lambda-\alpha)-\textrm{sign}(\lambda) ) Q(t,\lambda-\alpha)Q'(t,\lambda)+
\frac{m}{\alpha^2}Q(t,0)Q(t,\alpha),
\ee
where $Q(t,\lambda)$ is subjected to the condition 
\be
\label{e23}
\psi\equiv Q(t,\lambda-\alpha)+Q(t,\lambda+\alpha)\approx 0,
\ee     
and the Dirac bracket reads 
\be
\label{e24}
\{Q(t,\lambda),Q(t,\lambda ')\}=\frac{\alpha^2}{m}\sum_{k=-\infty}^{\infty}(-1)^k\delta(\lambda-\lambda '+(2k+1)\alpha).
\ee
The constraint $\psi$  can be explicitly solved:
\be
\label{e25}
Q(t,\lambda)=\sum_{k=-\infty}^\infty a_k(t)\psi_k(\lambda),
\ee
where
\be
\label{e26}
\psi_k=\frac{1}{\sqrt{2\alpha}} e^{\frac{i\pi}{2\alpha}(2k+1)\lambda} ,
\ee
form an orthonormal  and complete set in $L^2[-\alpha,\alpha]$.
Then the Hamiltonian  (\ref{e22}),  expressed in terms of the new variables, takes the form
\be
\label{e27}
H=\frac{\pi m}{4\alpha^3}\sum_{k=-\infty}^\infty(-1)^k(2k+1)a_ka_{-(k+1)},
\ee   
while the Dirac bracket   reads
\be
\label{e28}
\{a_k,a_n\}= \frac{i\alpha^2}{m}(-1)^k\delta_{k+n+1,0}.
\ee
After a simple redefinition of $a$'s 
\be
\label{e29}
a_k=\frac{\alpha}{\sqrt{m}}\left\{
\begin{array}{ll}
c_k&\textrm{ k-odd positive,}\\
c_{-(k+1)}&\textrm{ k-odd negative,}\\
\bar c_k&\textrm{ k-even  nonnegative,}\\
\bar c_{-(k+1)}&\textrm{ k-even  negative, }
\end{array}
\right.
\ee
we arrive at the infinite alternating  sum of oscillators with the odd frequencies
\be
\label{e29a}
H=\sum_{k=0}^\infty(-1)^k(2k+1)\frac{\pi}{2\alpha}\bar c_kc_k,
\ee  
with the standard brackets
\be
\label{e29b}
\{ c_k,\bar c_n\}=-i\delta_{kn}.
\ee
Thus the model (\ref{e20})  describes  the PU oscillator of infinite order with the odd frequencies. This fact  is interesting from the symmetry point of view because,  as it was shown  in  \cite{b5}, for the odd frequencies the symmetry group of the $N$-th order PU oscillator  is larger than for  the case of generic  frequencies;  it contains the conformal and  dilatation generators which together with the remaining symmetry generators (related to the time translation and the change of initial conditions) form the $(N-\frac 12)$-conformal non-relativistic  algebra.  Consequently, the question arises about the symmetry of the   nonlocal model (\ref{e20}).  
\section{Symmetries of the nonlocal PU model on the Lagrangian level}
 We start with the observation that the change of the initial conditions  produce new solutions. Due to the fact that any solution of the equation of motion  (\ref{e21}) is antiperiodic functions on the interval $[-\alpha,\alpha]$ we postulate the infinitesimal symmetry
\be
\label{e30}
t'=t,\quad q'=q+\delta q,  \quad \textrm { where } \delta q(t-\alpha)+\delta q(t+\alpha)=0.
\ee
Indeed,  taking 
\be
\label{e31}
\delta F=-\frac{m}{\alpha^2}\int_t^{t+\alpha}d\lambda\delta q(\lambda-\alpha)q(\lambda),
\ee    
one can check  that the  symmetry condition (\ref{e17})  holds.
The 
corresponding integral of motion reads
\be
\label{e32}
C=-\frac{m}{\alpha^2}\int_{-\alpha}^{\alpha}q(\lambda +t-\alpha)\delta q(\lambda +t).
\ee
Expanding $\delta q$ in the  Fourier serie (after substitution $q(t)\rightarrow q(t)e^{-\frac{i\pi t}{2\alpha}}$ )    we can identify the symmetry generators corresponding to the coefficients in the Fourier expansion
\be
\label{e33}
\delta q^c_k=\epsilon \cos(\frac{2k+1}{2}\omega t), \quad  \delta q^s_k=\epsilon \sin(\frac{2k+1}{2}\omega t), 
\ee     
where $\omega=\frac{\pi}{\alpha}$ and $k=0,1,2\ldots$ (for simplicity of the  further notation we can extend     $k$ to the negative  integers).  Consequently, the symmetry generators take the form 
\be
\label{e34}
\begin{split}
C^c_k&=-\frac{m}{\alpha^2}\int_{-\alpha}^\alpha q(\lambda+t-\alpha)\cos(\frac{2k+1}{2}\omega(\lambda+t)),\\
C^s_k&=-\frac{m}{\alpha^2}\int_{-\alpha}^\alpha q(\lambda+t-\alpha)\sin(\frac{2k+1}{2}\omega(\lambda+t)),
\end{split}
\ee
or in the differential realization  as commuting operators
\be
\label{e35}
\begin{split}
C_k^c=-\cos (\frac{2k+1}{2}\omega t)\frac{d}{dq}, \quad C_k^s=-\sin (\frac{2k+1}{2}\omega t)\frac{d}{dq}.
\end{split}
\ee
Note that with our convention  concerning $k$ (positive and negative) some generators are linearly dependent. 
\par
Now, the main  observation is that except the  quite natural symmetry (\ref{e30})  the  Lagrangian (\ref{e20})  possesses a more  interesting set of symmetries related to  the time transformation.
Namely, let us consider an arbitrary time transformation  $t'=t'(t)$ satisfying the condition
\be
\label{e36}
t'(t+\alpha)=t'(t)+\alpha,
\ee
and define the following transformation of the coordinate $q$
\be
\label{e37}
q'(t')=q(t)\left(\frac{dt'}{dt}\right)^{-\frac12}.
\ee 
Then one can check that the invariance condition (\ref{e16})   for the   nonlocal Lagrangian (\ref{e20})  is satisfied  (with $F=0$). In order to identify this symmetry let us take its infinitesimal form   
\be
\label{e38}
t'=t+\epsilon f(t), \quad  f(t+\alpha)=f(t).
\ee
Then 
\be
\label{e39}
\delta t=\epsilon f(t), \quad \delta_0 q=-\frac{\epsilon}{2}\dot f q;
\ee
due to  eq. (\ref{e18}) the corresponding integral of motion reads
\begin{align}
\label{e40}
C_f&=-\frac{m}{\alpha^2}q(t)q(t+\alpha)f(t)+\frac{m}{\alpha^2}\int_{0}^{\alpha} d\lambda q(\lambda+t-\alpha)\nonumber \\
&\left(\frac{1}{2} f'(\lambda+t)q(\lambda +t)+
 q'(\lambda+t)f (\lambda +t)\right) =\nonumber\\
&\frac{m}{2\alpha^2}\int_{-\alpha}^{\alpha}q(\lambda +t-\alpha) q'(\lambda+t) f(\lambda +t),
\end{align}
where in the second equality we have used (\ref{e21}).
Since the function $f$ is periodic with the period $\alpha$ it   can be expanded in terms of the functions
\be
\label{e41}
f_k^c(t)=\frac{1}{2\omega}\cos(2k\omega t), \quad  f_k^s(t)=\frac{1}{2\omega}\sin(2k\omega t) ,
\ee
for $k=0,1,2\ldots$ Substituting (\ref{e41})   into (\ref{e40})  we obtain a family of generators $L^c_k,L^s_k$ (as before  we may assume  $k$ integer, then $L^s_{-k}=-L^s_k$ and $L^c_{-k}=L^c_k$). One can easily find the  differential realizations of those generators:
\be
\label{e42}
\begin{split}
L_k^c= \frac{-1}{2 \omega} \cos(2k\omega t)\frac{d}{dt}-\frac{k}{2}\sin(2k\omega t)q\frac {d}{dq},\\
L_k^s= \frac{-1}{2 \omega} \sin(2k\omega t)\frac{d}{dt}+\frac{k}{2}\cos(2k\omega t)q\frac {d}{dq},\\
\end{split}
\ee
as well as the  commutation rules
\begin{align}
\label{e43}
[L_k^s,L_n^s]&=\frac{k-n}{2}L^s_{n+k}+\frac{n+k}{2}L^s_{n-k},\nonumber\\
[L_k^c,L_n^c]&=\frac{n-k}{2}L^s_{n+k}+\frac{n+k}{2}L^s_{n-k},\\
[L_k^s,L_n^c]&=\frac{k-n}{2}L^c_{n+k}+\frac{n+k}{2}L^c_{n-k}.\nonumber
\end{align}
Moreover,  the commutators of $L$'s  with $C$'s read 
\be
\label{e43a}
\begin{split}
[C_k^s,L_n^s]&=\frac18\left ((2n-2k-1)C^s_{k-2n}+(2n+2k+1)C^s_{k+2n}\right),\\
[C_k^s,L_n^c]&=\frac18 \left((-2n+2k+1)C^c_{k-2n}+(2n+2k+1)C^c_{k+2n}\right),\\
[C_k^c,L_n^s]&=\frac18 \left((2n-2k-1)C^c_{k-2n}+(2n+2k+1)C^c_{k+2n}\right),\\
[C_k^c,L_n^c]&=\frac18 \left((2n-2k-1)C^s_{k-2n}-(2n+2k+1)C^s_{k+2n}\right).
\end{split}
\ee
Summarizing,  the symmetry algebra  is defined by (\ref{e43}) and (\ref{e43a});  as usual on the Lagrangian level  there  is no central change.
\par 
In order to have a better insight  into  this  algebra let us note that the generators  $L_0^c,L^c_1,L^s_1$ form a three-dimensional subalgebra:
\begin{align}
\label{e44}
[L_0^c,L_1^c]=L_1^s,\quad 
[L_1^s,L_0^c]=L_1^c,\quad 
[L_1^s,L_1^c]=L_0^c,
\end{align}
in which, after the substitution,
\be
\label{e45}
\tilde H=-(L^c_1+L_0^c),\quad K=L^c_1-L_0^c,\quad D=L_1^s,
\ee
we recognize the $sl(2,R)$ algebra  
\be 
\label{e46}
[D,\tilde H] =\tilde H,\quad [K,D]=K,\quad [\tilde H,K]=-2D.
\ee
Now,   let us remind  that 
the free classical motion (in general, in the sense of  higher derivatives theory) possesses the conformal symmetry (i.e., $SL(2,\mR)$ symmetry, see \cite{b6g}-\cite{b6i}) which acts on time variable $\tilde t$ according to the formula
\be
\label{e47}
\tilde{ t}'=\frac{ a \tilde  t+b}{c \tilde t+d }, \qquad \quad  
\left(
\begin{array}{cc}
a&b\\
c&d
\end{array}
\right)\in SL(2,\mR).
\ee 
This symmetry can be transformed to the harmonic oscillator case   (the PU oscillator with the odd frequencies in the case of higher derivatives)  by means of the  	 Niederer transformation (or its generalization to the higher derivatives, see \cite{b4h,b5} for further references)  in which   time  of the free motion $\tilde t$  is related 
to   "oscillator" time  $t$ by the formula	
\be
\label{e48}
\tilde t=\tan(\omega t).
\ee   
Since our model is a natural extension of the harmonic oscillator  we can try to  induce directly the  symmetry from the action
of $SL(2,\mR)$ on $\tilde t $. Indeed, taking the standard realization of the conformal generators
\be
\label{e49}
\tilde H=-\sigma_+,\quad D=\frac{1}{2}\sigma_3,\quad K=\sigma _-,
\ee 
 we obtain, by virtue of eq. (\ref{e47}),   the following infinitesimal action on $t$ 
\begin{align}
\label{e50} 
\tilde H: \quad t'&=t-\frac{\epsilon}{2\omega}(\cos(2\omega t)+1),\nonumber\\
 K: \quad t'&=t+\frac{\epsilon}{2\omega}(\cos(2\omega t)-1),\\
 D: \quad t'&=t+\frac{\epsilon}{2\omega}\sin(2\omega t)\nonumber,
\end{align}
which perfectly agrees with the functions $f^c_0,f^c_1,f^s_1$ and  the relation (\ref{e45}). 
\par
Moreover, 	the Hamiltonian (\ref{e22}) coincides with the   generator of the time translation, i.e.,   $L_0^c=-\frac{1}{2\omega}H$ which  again agrees with the fact that  the dynamics of the  considered model is related   to the different choice of the basis of the conformal algebra  $H=\omega(\tilde H+K)$ (as in the case of the  harmonic  and PU oscillators, see \cite{b4h,b7h} and the references  therein) 
\par
We see that $L$'s define an  infinite-dimensional extension   of the $sl(2,\mR)$ algebra.  Things become more transparent  if  we use the functions 
$f_k(t)=\frac{1}{2\omega}e^{2ik\omega t}$ with  $k$-integer,   instead of the functions  (\ref{e41}),  and  $\psi_k$ (eq. (\ref{e26})), with $k$ integer instead of  the functions appearing in (\ref{e33}). Then the corresponding generators   $L_k$ and $C_k$ (in the complexified  algebra)   can be expressed in the form
\be
\label{e51}
L_k=L_k^c+iL^s_k, \quad C_k=\frac{1}{\sqrt{2\alpha}}(C_k^c+iC_k^s), \quad k-\textrm{integer},
 \ee  
and satisfy the following commutation relations
\begin{align}
\label{e52}
[L_k,L_n]&=i(k-n)L_{k+n},\\
\label{e52a}
[C_k,L_n]&=\frac{i}{4}(2k+1+2n)C_{k+2n}.
\end{align} 
Form eq. (\ref{e52a}) we infer that the symmetry algebra can be split in two parts spanned by  $C_k$'s with $k$ even or odd, respectively; thus we define  
\be
C^+_k=C_{2k}, \quad C^-_k=C_{2k-1}, \quad k - \textrm{integer}.
\ee
As a result the commutation rules  (\ref{e52a})  are converted into  the following ones
\be
[C^\pm_k,L_n]=i(k+\frac n2 \pm\frac14)C_{k+n}^\pm,
\ee
in which we recognize two  representations of the Virasoro algebra acting on $C^\pm$'s (with the nonhomogeneous part  $\pm\frac14$, see, e.g., \cite{b8}). At this point, let us recall that the  PU oscillator  with the odd frequencies enjoys the $(N-\frac 12)$-conformal non-relativistic symmetry  which, for fixed $N$, possesses   infinite-dimensional extensions (see, \cite{b7b}-\cite{b7i}). 
However,  there is no direct limit in these algebras  as $N$ tends to   infinity (in contrast to  the Lagrangian or the  Hamiltonian of the PU oscillator).  They   differ from the one  considered  above    by  the adjoint  action of the Virasoro generators.
\par 
Finally, let us note that  the model (\ref{e20}) as well as  our former considerations can be extended to  the case of coordinates taking their values in three-dimensional space $\vec q=(q^\alpha)$ (or, in generally, in $d$-dimensional one). Then the symmetry generators (\ref{e35}) and (\ref{e42})   extend naturally  to the  vector case: $C^\alpha$'s and $L$'s, respectively. Moreover,  we have an  additional family of  generators 
 \be
\label{e54}
J_{k}^{\alpha c}=f^c_kJ^\alpha, \quad J_{ k}^{\alpha s}=f^s_kJ^\alpha, 
\ee 
related to the symmetry 
\be
\label{e53}
t'=t, \quad \vec q'=R(t)\vec q,
\ee  
where $R$ is a  time-depended rotation satisfying  $R(t+\alpha)=R(t)$. 
As earlier (see (\ref{e51}))  it is instructive to pass to the complexification  of the algebra and define  the new  generators  $J^\alpha_k=J_k^{\alpha c}+iJ_k^{\alpha s}$. Then the  commutation rules containing $J$'s  take the  Kac-Moody  form 
\begin{align}
\label{e53a}
[J_{k}^\alpha,J_{n}^\beta]&=\frac{1}{2\omega}\epsilon^{\alpha\beta\gamma}J^\gamma_{ k+n},\nonumber \\
[J_k^\alpha,C_n^{\beta\pm}]&=\frac{1}{2\omega}\epsilon^{\alpha\beta\gamma}C^{\gamma\pm}_{n+k},\\
[J^\alpha_k,L_n]&=ik J^\alpha_{k+n},\nonumber
\end{align}
which coincide with the ones described  in refs. \cite{b7b}-\cite{b7h} for a finite order conformal algebras (the difference is in the action of the Virasoro subalgebra).   
\section{Symmetry realizations on the Hamiltonian and quantum  levels}
 In this section we discuss  the  symmetry algebra in  Hamiltonian  approaches and show that  it forms a natural central extension of the algebra on the Lagrangian level; on the quantum level the central extension appears also for  the Virasoro algebra. 
\subsection{Constrained approach}
First let us note that the Hamiltonian (\ref{e22}) as well as  the symmetry generators $C$'s and $L$'s can be rewritten, due to the correspondence (\ref{e6}),  as follows
\be
\label{e56}
\begin{split}
H&=-\frac{m}{2\alpha^2}\int_{-\alpha}^\alpha d\lambda Q(t,\lambda -\alpha)Q'(t,\lambda),\\
L_k^c&=\frac{m\omega}{4\pi^2}\int_{-\alpha}^{\alpha}d \lambda Q(t,\lambda-\alpha) Q'(t,\lambda) \cos(2\omega k(\lambda +t)),\\
L_k^s&=\frac{m\omega}{4\pi^2}\int_{-\alpha}^{\alpha}d \lambda Q(t,\lambda-\alpha) Q'(t,\lambda) \sin(2\omega k(\lambda +t)),\\
C_k^c&=-\frac{m}{\alpha^2}\int_{-\alpha}^{\alpha}d \lambda Q(t,\lambda -\alpha)\cos\left(\frac{2k+1}{2}\omega(\lambda +t)\right),\\
C_k^s&=-\frac{m}{\alpha^2}\int_{-\alpha}^{\alpha}d\lambda Q(t,\lambda -\alpha)\sin\left(\frac{2k+1}{2}\omega(\lambda +t)\right).
\end{split}
\ee
Now, using the  functional Dirac bracket (\ref{e24}) we can find all commutation rules and consequently the symmetry algebra on the Hamiltonian level. The computations are straightforward but rather tedious, due to the fact  that we have to  take carefully  into account  the support of each delta function appearing  in the Dirac bracket (\ref{e24}). As a consequence, we obtain almost the same relations on the Hamiltonian  level as on the Lagrangian one -- the only difference appears in commutators between  $C$'s; in this case we obtain a central extension, namely
\be
\label{e57}
[C_k^c,C^s_n]=\frac{m}{\alpha}(-1)^{n-1}\delta_{nk}, \quad k,n=1,2,3\ldots
\ee 
\subsection{Unconstrained approach} 
As we saw in Section 3 solving the constraint (\ref{e23}) on the Hamiltonian level allow us to  identify our model with the  PU oscillator of  the infinite order (see (\ref{e29a})). Therefore, it is instructive to find the form of  the symmetry generators in terms of classical counterparts  the creation and annihilation operators  $c$'s  (equivalently $a$'s ). We will see that the  symmetry generators (except the Hamiltonian) are not   direct extensions of the ones for the classical  PU oscillator  \cite{b4g}.
In order to do  this we can  express, by virtue of. eq. (\ref{e25}),   the  symmetry generators in terms  of $a$'s.  After some troublesome  calculations we arrive at the following form of the generators 
\be
\label{e58}
\begin{split}
C_k^c&=\sqrt{\frac{\alpha}{2}}(C_k+\bar  C_k),\quad C_k^s=i\sqrt{\frac{\alpha}{2}}(\bar C_{k}-C_k), \\
L_k^c&=\frac{1}{2}(L_k+\bar L_k), \quad L_k^s=\frac{i}{2}(\bar L_k-L_k),
\end{split}
\ee
where
\be
\label{e59}
C_k=\frac{m}{\alpha^2}e^{\frac{i\pi}{2\alpha}(2k+1)(t-\alpha)}\bar a _k,
\ee
and 
\be
\label{e60}
\begin{split}
L_n&=-\frac{m}{8\alpha^2}e^{2in\omega t}\sum_{k=-\infty}^\infty(-1)^k (2k+1)a_{-k-2n-1}a_k\\
&=\frac{m}{4\alpha^2}(-1)^{n+1}e^{2in\omega t}\sum_{k=0}^\infty (-1)^k(2k+1)\bar a_{k+n}a_{k-n}. 
\end{split}
\ee 
 The following identity  appears to be  useful
 \be
 \label{e61}
 \bar a_k=a_{-(k+1)}.
 \ee
Now,  using the  bracket  (\ref{e28}) we can confirm the commutation relations (\ref{e43}), (\ref{e43a}) and (\ref{e57}) obtained within the  Hamiltonian framework with constraints.  The calculations  are straightforward but rather long so we only mention that,  using (\ref{e28}) and (\ref{e61}),   one can check the relation (\ref{e52}) which, in turn, easily implies (\ref{e43}).
\par 
Of course, one can express,  by virtue  of (\ref{e29}),   the generators (\ref{e59}) and (\ref{e60}) in terms of $c's$ (for the Hamiltonian  see   (\ref{e29a})).
Consequently we obtain a non-standard realization  of the Virasoro algebra (\ref{e52})  what will be reflected  in the form of the  central extension  on  the quantum level  (see (\ref{e66}) or (\ref{e68})). 
 Moreover,  let us note  that  
 \be
 \label{e62}
 \bar C_n=C_{-(n+1)},\quad \bar L_n=L_{-n}\quad L_0=\bar L_0=L_0^c=-\frac{1}{2\omega}H.
 \ee
 \par
 Finally, the generalization of  our considerations on the Hamiltonian  level  to  the   three-dimensional  (in general $d$-dimensional )  case  $Q^\alpha $  is   straightforward  and the main difference is that we have  additional generators  corresponding  to the rotational symmetry  (cf.  eqs. (\ref{e54}) and (\ref{e53}))
\be
\label{e63}
\begin{split}
J_n^\alpha&=\frac{m\omega}{4\pi^2}\epsilon^{\alpha\beta\gamma}\int_{-\alpha}^{\alpha }d \lambda Q^\beta(t,\lambda-\alpha)Q^\gamma(t,\lambda)e^{2i\omega n(t+\lambda)}\\
&=\frac{-im\omega}{4\pi^2}\epsilon^{\alpha\beta\gamma}e^{2i\omega tn}\sum_{k=-\infty}^\infty(-1)^k a^\gamma_{-k-2n-1} a^\beta_k,
\end{split}
\ee
which satisfy  the commutation rules in  (\ref{e53a}).
\subsection{Quantum level}
In this section we  analyse symmetries of the nonlocal PU model on the quantum level with special emphasis put on the Virasoro algebra. Following the standard reasoning we replace the  classical  variables  $c_k,\bar c_k$, satisfying (\ref{e29b}),    by the creation and annihilation operators    $\hat c_k, \hat c_k^+$ 
\be
\label{e64}
[\hat c_k,\hat c_n^+]=\delta_{kn},
\ee 
for $k,n=0,1,2\ldots$  Next,  using the quantum counterpart of the relation (\ref{e29})   we   define the  operators $\hat a_k,\hat  a_{-(k+1)}=\hat a^+_k$.  Then   we have the following commutator rules   
\be
\label{e65}
[\hat a_k,\hat a_n]=\frac{-\alpha^2}{m}(-1)^k\delta_{k+n+1,0}.
\ee
Now, we can translate the symmetry  generators to the quantum level. In the case of  $C$'s,  $J$'s and $L_n$ with $n\neq 0$ it can be done directly  (there is no problem with  ordering of  the operators, see eqs. (\ref{e59}), (\ref{e60}) and (\ref{e63})); however,  the generator $L_0$  requires more careful analysis.   To this end let us compute the following commutator $[\hat L_n, \hat L_{-n}]$ for $n\neq 0$. Using the   second form in (\ref{e60})   and (\ref{e65})     we arrive, after some computations, at the following result 
\be
\label{e66}
[\hat L_n,\hat L_{-n}]=  -2n\hat L_0+\frac {1}{12}(n^3-\frac{1}{4}n),
\ee 
where
\be
\label{e67}
\hat L_0=\frac{-m}{4\alpha^2}\sum_{k=0}^\infty(-1)^k(2k+1)\hat a_k^+\hat a_k.
\ee
Now let us express $\hat L_0$ in terms of the creation and annihilation operators
\be
\label{e69}
\hat L_0=\frac{1}{4}\sum_{\substack{k=0\\ k\textrm{-odd}}}^\infty (2k+1)\hat c_k^+\hat c_k-\frac 14\sum_{\substack{k=0\\ k\textrm{-even} }}^\infty (2k+1)\hat c_k\hat c_k^+.
\ee 
Performing the  regularization procedure by mens of the Riemann zeta function $\hat L_0$ becomes   a normally ordered  operator  (proportional to the quantum  Hamiltonian)   and the following  commutation rules
\be
\label{e68}
[\hat L_n,\hat L_k] =(k-n)\hat L_{n+k}+\frac {1}{12}(n^3+\frac{3}{4}n)\delta_{k+n,0},
\ee
are satisfied; in the case of $d$ dimension  there is  the additional  factor $d$ in front of the  central charge in (\ref{e68}).
Finally, by adding a constant to the normally ordered $\hat L_0$  and rescaling $\hat L_n \rightarrow -\hat L_n$ one obtains the   standard form  of the   centrally extended Virasoro algebra.
\section{Conclusions}
We have discussed the symmetry properties of a simple nonlocal theory which provides  the extension, to the  case of an   infinite number of degrees of freedom, of the PU oscillator with frequencies proportional  to consecutive  odd integers.  Since  the PU  oscillator  of order $N$  with such frequencies enjoys $(N-\frac 12)$-conformal non-relativistic symmetry   the resulting  symmetry  algebra  can be viewed as an extension  of the  latter one to  infinite order $N=\infty$.  It can be described within the framework of the Noether theorem  both in its Lagrangian and Hamiltonian forms:  
the original $sl(2,\mR)$ subalgebra  of  conformal non-relativistic algebras  is extended  to the Virasoro algebra; the additional charges (outside the Virasoro subalgebra) form two  representations of the latter under its  adjoint action. They commute with each other on the Lagrangian level while on the Hamiltonian one we are  dealing with  the  central extension  defined by eq. (\ref{e57}).   Moreover,   on the quantum level  the Virasoro algebra is also centrally extended (eq. (\ref{e66}) or (\ref{e68})). 
\par
Let us note that the model described here resembles to some extent the string model. In the Hamiltonian framework we are dealing with string obeying antiperiodic boundary conditions (cf. eq. (\ref{e23})) with the dynamics restricted to left-movers only (eq. (\ref{e4})). There are, however, some important differences. First, $Q$'s (or $q$'s) take their values in Euclidean space  and there are no  additional constraints on them.
 Moreover,  the energy is the alternating sum of independent  oscillators energies (eq. (\ref{e29a})) which makes the system unstable and alters the vacuum energy.  The model, although basically non-interacting, exhibits interesting structure.  It is worthy of further study in several directions. First, one can  consider the supersymmetric extensions  following the construction proposed for finite order \cite{b4i,b4ll}. Second, the above  mentioned   problem of (un)stability of the system (consequently  the form of $\hat L_0$)  might be studied  along the lines proposed in ref.  \cite{b4} for the quartic PU oscillator.  According to this reasoning the unstability of the model could be cured at the expense of introducing negative  norm states. Also  the relation of our symmetry algebra to those considered in refs. \cite{b7b}-\cite{b7i} as well as  a non-relativistic version of AdS$\backslash $CFT correspondence for the discussed model are  interesting questions. 
\par
{\bf Acknowledgments.}
Special thanks are to Piotr Kosi\' nski who suggested the problem studied here.  The discussions with Joanna  and Cezary  Gonera as well as  Pawe\l\  Ma\'slanka are gratefully acknowledged. The research was supported by the grant of National Science  Center number DEC-2013/09/B/ST2/02205.
  
\end{document}